\documentclass[%
reprint, 
superscriptaddress,
 amsmath,amssymb,
 aps,
pra,
]{revtex4-2}

\usepackage{graphicx}
\usepackage{dcolumn}
\usepackage{bm}

\usepackage[utf8]{inputenc}
\usepackage{makecell}
\usepackage{hyperref}
\usepackage{mathrsfs}
\usepackage{graphicx} 
\usepackage{siunitx}

\begin{document}

\preprint{Simonaitis/Low-Energy Spectroscopy}

\title{Electron Energy Loss Spectroscopy of 2D Materials in  a Scanning Electron Microscope}

\author{John W. Simonaitis}
 \email{johnsimo@mit.edu}
 \affiliation{Research Laboratory of Electronics, Massachusetts Institute of Technology, Cambridge, Massachusetts 02139, USA}

\author{Joseph A. Alongi}%
 \affiliation{Research Laboratory of Electronics, Massachusetts Institute of Technology, Cambridge, Massachusetts 02139, USA}

\author{Benjamin Slayton}%
\affiliation{%
Department of Electrical and Computer Engineering. University of California, Davis, California 95616, USA
}%

\author{William P. Putnam}
 \affiliation{%
 Department of Electrical and Computer Engineering. University of California, Davis, California 95616, USA
}%

\author{Karl K. Berggren}
 \affiliation{Research Laboratory of Electronics, Massachusetts Institute of Technology, Cambridge, Massachusetts 02139, USA}

\author{Phillip D. Keathley}
 \email{pdkeat2@mit.edu}
 \affiliation{Research Laboratory of Electronics, Massachusetts Institute of Technology, Cambridge, Massachusetts 02139, USA}

\date{\today}

\begin{abstract}

This work demonstrates electron energy loss spectroscopy  of 2D materials in a 1-30 keV electron microscope, observing 50-times stronger electron-matter coupling relative to 125~keV microscopes. We observe that the universal curve relating beam energy to scattering holds for the transition from bulk graphite to graphene, albeit with a scale factor. We calculate that optimal coupling for most 2D materials and optical nanostructures falls in this range, concluding that spectroscopy of such systems will greatly benefit from use of this previously unexplored energy regime.

\end{abstract}

\maketitle

In recent years there has been explosive growth in the study of two-dimensional (2D) materials, ushered in by the discovery of the remarkable optical, mechanical, and electronic properties of graphene \cite{novoselov_electric_2004}. Since then, many other 2D materials have been discovered with a plethora of exemplary properties. However despite this promise, creating practical devices with such materials has proven elusive, due in part to their extreme surface sensitivity and the lack of pristine large-area samples.

Electron energy loss spectroscopy (EELS) has proven to be a powerful technique for studying 2D materials due to its nanometer spatial resolution, which allows for direct probing of the optical and electronic structures of small pristine regions \cite{eberlein_plasmon_2008}, and ability to probe defects and dark modes inaccessible by other proves \cite{garcia_de_abajo_optical_2021}. If ultrafast triggering of the electron beam is used, femtosecond resolution of dynamics in these materials also becomes possible, albeit at the cost of a significant reduction of the beam current \cite{hu_indirect_2023, cremons_femtosecond_2016, danz_ultrafast_2021, kurman_spatiotemporal_2021}.

For such experiments, transmission electron microscopes (TEMs) equipped with commercially available electron spectrometers are most commonly used. This makes sense as historically important materials, such as alloys and semiconductors, are readily analyzed by TEMs if thinned. The optimal thickness for such experiments is generally selected to be in the 30-100~nm range, with the exact value dependent on the sample composition and beam energy, and the goal being to maximize the probability of a single electron-material interaction while minimizing multiple scatterings that distort or fully absorb the electron. Since these specimens are thick enough to have thickness-invariant properties, this thinning has little effect on the resulting spectrum.


However, when TEMs are used to study 2D materials or photonic nanostructures, the specimen thickness can not be adjusted without fundamentally changing the properties of the material. Due to this constraint, inelastic scattering for 2D materials is extremely weak, with loss peaks containing less than 1$\%$ the intensity of the beam \cite{eberlein_plasmon_2008, kurman_spatiotemporal_2021, persichetti_folding_2011}. For fine spectral features, this weak signal can easily be washed out by the zero-loss peak (ZLP) energy spread and noise. When ultrafast pump-probe experiments are conducted, the additional four-order-of-magnitude loss of beam current exacerbates this issue \cite{arbouet_ultrafast_2018}, meaning that ultrafast electron spectroscopy of such specimens is difficult and thus rarely done. 

This weak TEM interaction is part of the reason for which low-energy electron microscope (LEEM) systems are used for 2D materials spectroscopy \cite{geelen_nonuniversal_2019, jobst_quantifying_2016}. However these tools, which operate in the 1-100~eV energy range, have stringent vacuum requirements, high sensitivity to vibrations/magnetic fields, and limited ultimate resolution, meaning that LEEMs are far less common than SEMs and TEMs. Despite this, they remain important tools for understanding 2D materials, with several recent high-profile studies on the band structure and energy-scaling behavior of 2D materials \cite{geelen_nonuniversal_2019, jobst_quantifying_2016}.

Scanning electron microscopes (SEMs) potentially offer a compromise between LEEMs and TEMs. Since they have significantly lower beam energies (and thus velocities) than TEMs, the beam spends more time interacting with the sample, leading to potentially stronger interactions. Unlike LEEMs, they are widely available and offer robust, few-nm imaging with moderate vacuum and shielding requirements. Their wide dynamic range is also appealing: SEMs can operate from 300~eV to 30~keV corresponding to approximately 3$\%$ to  $33\%$ the speed of light, leading to a high likelihood of phase-matching and optimal coupling.

However, SEM-based EELS is rare -- in part due to the lack of commercially available instrumentation at such energies, as well as the historical lack of samples thin enough to transmit these energies.  Only with the growing interest in 2D materials and photon-induced near-field electron microscopy (PINEM) of optical nanostructures has such apparatus become worthwhile to construct.  To our knowledge, only one other group has demonstrated a sub-eV resolution SEM-based spectrometer ~\cite{shiloh_quantum-coherent_2022}. In this work, they demonstrated PINEM spectroscopy of a laser-excited nanometric tip with 10.4 keV electrons. However, they did not study other systems, such as 2D materials, and to our knowledge no other group has the capability to do measurements in this energy range. 

In this work, we explore the first use of a 1-30~keV SEM to perform EELS analysis 2D materials and optical nanostructures. After constructing and calibrating our spectrometer, we use it to measure the loss spectrum of graphene, observing loss spectra two-orders-of-magnitude stronger than experiments in TEMs. We adapt a bulk model for the inelastic mean free path (IMFP) versus beam energy for use in 2D materials and show that the universal IMFP curve for graphite still holds for graphene, albeit with 5-times stronger coupling than expected. From this, we observe that the optimal coupling range for graphene EELS falls in the energy range of our instrument. We expand this analysis to a wide range of 2D materials, from 10~eV to 300~keV, and experimentally show that the optimal electron energy for analyzing most 2D materials occurs at SEM energies, and that order-of-magnitude signal improvements over TEMs can be widely expected, given present technical limitations are overcome. 

To see if this enhancement applies to other systems with weak coupling, we simulate a PINEM interaction with optical nanostructures, concluding that optimal coupling occurs at SEM energies for a wide range of wavelengths and structure sizes. Together, our simulations and experiments lead us to conclude that this previously unexplored 1-30~keV electron spectroscopy range offers greatly improved performance for a range of emerging low-dimensional materials and nanostructures of interest, and will enable ultrafast electron pump-probe experiments that were previously impossible.
 
\begin{figure}[t]
    \centering
    \includegraphics[width=\linewidth]{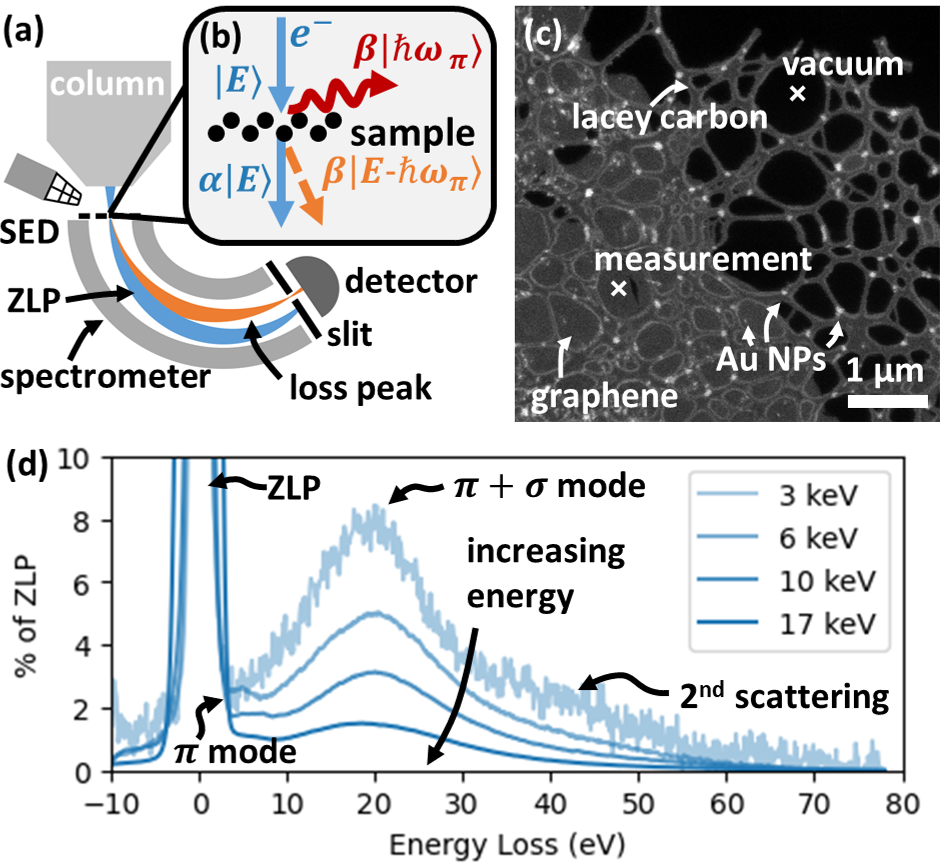}
    \vspace{-8mm}
    \caption{Rendering of our experiment. (a) Schematic of our SEM column, secondary electron detector (SED), electron spectrometer, and sample. After interaction, the spectrometer plates deflect each electron proportional to its energy, which is then filtered by a slit and detected. (b) Rendering of our sample plane, in which our specimen (shown in Fig. 1c) generates an excitation in the material with likelihood $|\beta|^2$.  This results in an attenuated probability of measuring an initial energy electron (our ZLP) with probability $|\alpha|^2$,  and a $|\beta|^2$ chance of the electron losing $\hbar\omega_\pi$ energy.  (c) SEM micrograph of our sample, which consists of graphene on lacey carbon. (d) Graphene loss spectrum at various energies, showing an increase in the loss peak as the energy is decreased.}
    \label{fig:exp_overview}
    \vspace{-3mm}
\end{figure}


An overview of our experiment is shown in Fig. 1a. In it, a custom-built cylindrical 127\unit{\degree} electrostatic analyzer \cite{roy_design_1990, bryce_127_1973}  has been integrated into the stage of a modified FEI XL-30 SEM. When the electron loses energy interacting with the sample (Fig. 1b), it spends a longer time between the plates, leading to a stronger electrostatic deflection and spatial separation, which is filtered by a 5~\unit{\micro\meter} slit and detected. This apparatus achieves 0.7~eV energy resolution at 10~keV, as measured by the full width at half maximum (FWHM) of the ZLP. The details of the spectrometer alignment, operation, and resolution measurements are included in the supplement.

We then place a TEM grid with graphene on lacey carbon into the input aperture plane. An SEM micrograph of this is shown in Fig. 1c, and examples of a measurements over vacuum and through graphene are illustrated. These graphene measurements result in the spectrum shown in Fig. 1d, taken for a range of beam energies. In it, both of the so-called  $\pi$ and $\pi + \sigma$  modes are visible at 5~eV and 20~eV respectively \cite{eberlein_plasmon_2008, wachsmuth_plasmon_2014}, clearly distinguishing it as graphene. The peak of the ZLP has been normalized to 100\%, meaning that the peak of the $\pi + \sigma$ mode is at around 8\%, a significant fraction of the ZLP. Integrating this 3 keV loss peak we find it contains roughly 30\% of the total current, which is over an order-of-magnitude stronger than that of the equivalent TEM measurement \cite{persichetti_folding_2011}. Importantly, at this low energy a second peak becomes visible at 40~eV, indicating strong enough coupling for two scattering events to occur.

From this ratio of the loss peak current to the total measured current, we estimate a quantity known as the electron inelastic mean free path (IMFP), which quantifies the probability of an energy loss event (the goal of EELS) as a function of the sample thickness. We denote this as $\lambda_i$, and can calculate it using the Log-Ratio method \cite{malis_eels_1988}, given by
 \begin{equation}
     \frac{t}{\lambda_i} = \ln(\frac{I_t}{I_0}),
 \end{equation}
 where $I_t$ is the integral of the full spectrum, $I_0$ the integral of just the ZLP, and $t$ the specimen thickness, which is a monolayer. For the purposes of this work, we will always use the thickness-to-IMFP ratio ($t/\lambda_i$) as a proxy for the strength of the electron-matter interaction. We set the threshold for the ZLP integral by measuring the width in which 95\% of the beam is captured when over vacuum, and integrating just this region. 

\begin{figure}[t]
    \centering
    \includegraphics[width=\linewidth]{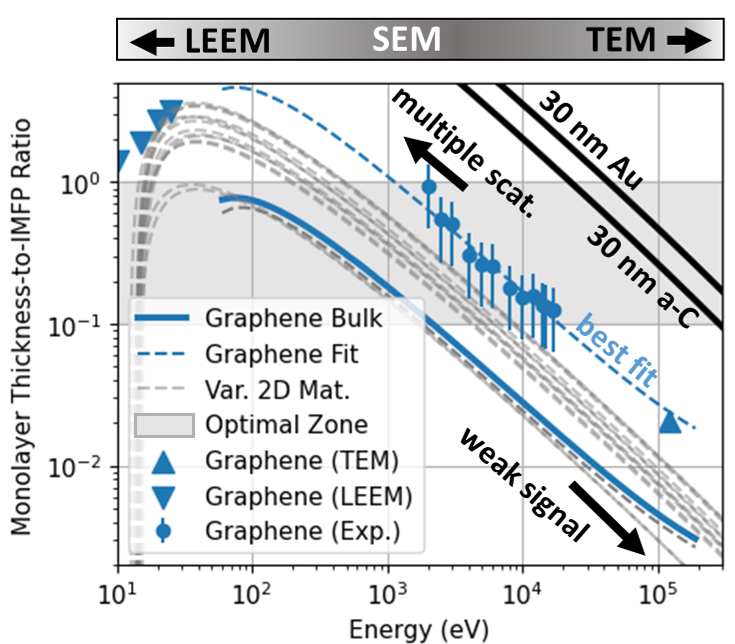}
    \vspace{-8mm}
    \caption{(a) Calculation of the thickness to IMFP ($t/\lambda_i$) versus energy for a variety of 2D materials. The graphene bulk curve was adapted from Jablonski et al. \cite{jablonski_calculations_2023}, and is based on optical data. The gold, amorphous carbon, and various other 2D material curves were estimated using the  Bethe approximation, with details in the supplement. We plot measurements of the IMFP for LEEM \cite{geelen_nonuniversal_2019} and TEM \cite{persichetti_folding_2011} energies from the literature with triangles. Our measurements are the blue circles with error bars. While the slope is consistent, the coupling measured in all of these experiments is approximately five times stronger than predicted. }
    \label{fig:IMFP_calcs}
    \vspace{-3mm}

\end{figure}


We plot the thickness-to-IMFP ratio ($t/\lambda_i$) for a variety of materials in Fig. \ref{fig:IMFP_calcs}a. The horizontal axis corresponds to the incident electron energy, and the top labels show roughly where LEEMs, SEMs, and TEMs operate. The optimal coupling region in is shaded in grey, and occurs for a $t/\lambda_i$ values between 0.1 to 1. If this ratio is greater than 1, it means that on average $m \geq 1$, and so more than one scattering event will occur per electron, complicating analysis. If this ratio gets too large, most electrons will be absorbed resulting in little to no signal. If the ratio is too small, very few electrons will be scattered, and the resulting spectrum obscured.  

 For most samples, this optimal $t/\lambda_i$ range is achieved by thinning the sample to around 30-100~nm. This is clearly shown by the black curves (representing 30~nm of gold (Au) and amorphous carbon (a-C)) in the top right of Fig. \ref{fig:IMFP_calcs}a intersecting this zone at TEM energies. These were estimated using the modified Bethe equation, which has been well-established in bulk materials for decades \cite{jablonski_calculations_2023}, and is given by
 
\begin{equation}
\label{eqn:bethe}
    \lambda_i(E_0)=\frac{\alpha(E_0) E_0}{E_p^2[\beta \ln (E_0)-\frac{C}{E_0}+\frac{D}{E^2}]},
\end{equation} 
where $\alpha$ is a relativistic factor, $E_p$ the plasmon energy, and $\gamma$, $C$, $D$, and $U$ empirical fitting parameters taken from Jablonski et al. \cite{jablonski_calculations_2023}. These are derived from the material density, $\rho$, the number of valence electrons, $N_v$, and the molar mass of the constituent atoms/molecules. Their exact calculation is given in the supplement. 

A priori, it is unclear if this model will work for 2D materials as such specimens are dominated by surface rather than the bulk effects assumed in Bethe model. In fact it has recently been shown that for LEEM measurements of graphene, the Bethe model fails entirely \cite{geelen_nonuniversal_2019}. However, we can still gain significant insight from assuming these simplified bulk-verified models scale linearly with thickness to the 2D limit.  The solid  blue curve shows this for graphene, and made using the full Penn algorithm, which is more accurate than the Bethe model \cite{penn_electron_1987} and estimates the IMFP from optical data. This was done for a variety of materials by Jablonski et al. \cite{jablonski_calculations_2023}, including graphite, which is plotted with the blue dashed line in Fig. 2a. 

From experiments in the literature \cite{persichetti_folding_2011} it is clear that for TEMs this thickness-to-IMFP ratio both substantially deviates from the Penn model and falls far below the optimal coupling range. This is seen by the upright triangle in the bottom right corner of Fig. 2a. As the beam energy is  reduced we expect the IMFP to decrease due to greater sample interaction, meaning that the monolayer-thickness-to-IMFP ratio should increase until a peak value which for bulk samples generally around 100~eV. We can see the IMFP-ratio  past this peak with the LEEM data from \cite{geelen_nonuniversal_2019} (the triangles in the top left), ratios greater than 1, indicating strong scattering.  From these prior measurements, we can infer that the optimal free-electron coupling to graphene must occur for some energy between these extremes. However this measurement has never been done due to the lack EELS in this range, and so the exact IMFP energy dependence has never been observed.


This lack of data led to our SEM-based measurements, resulting in the blue circles plotted. These points assume monolayer graphene (Ted Pella 21710) with a thickness of 0.335~nm. The error bars are due to uncertainty in the graphene thickness, which we assume to be from 0.7 to 2 layers thick, corresponding to an assumption of defects in the graphene of up to 30\%, or up to an extra monolayer of graphene or amorphous carbon contamination. In order to reduce the effects of contamination (and an overestimation of the IMFP), we moved the probe for each measurement. If the same point was used repeatedly, the sample became visibly thicker over time, leading to the appearance of stronger coupling. Various sample spectra at different energies are shown in Fig. 2b, which highlight how the loss peak intensity increases as the energy decreases. If we fit our measurements to the Penn model, allowing for variation in the assumed thickness, we find excellent agreement when the graphene is effectively 5-times thicker, as seen by the blue dashed curve. This agrees well with the TEM measurements, as well the LEEM measurements, if we linearly extrapolated the Penn model.


In order to see if this result applies to other 2D materials,  we calculated the IMFPs for various of 2D materials (Var. 2D Mat.) to give a general sense of where these curves should lie relative to graphene. These are shown by the grey dashed curves. Nearly all other materials are expected to have stronger scattering than graphene due to its low atomic number, as well as the fact that they are thicker. For this same reason we hypothesize that the effective thickness scaling will be less, since for thicker specimens surface effects will dominate less.



\begin{figure}[t]
    \centering
    \includegraphics[width=\linewidth]{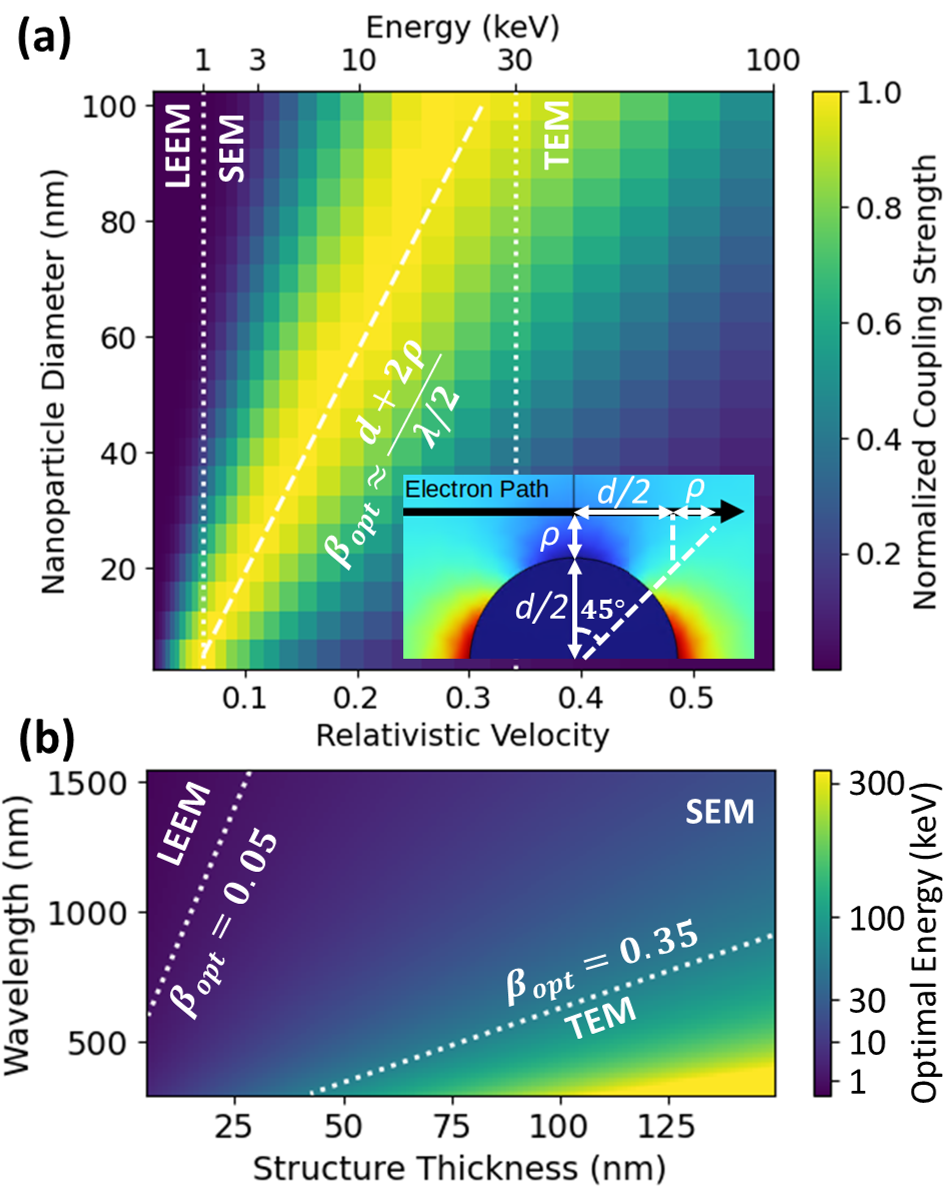}
    \vspace{-8mm}
    \caption{Simulation of an electron passing distance $\rho$ from a gold nanoparticle of diameter $d$. (a) Plot of the coupling factor versus electron velocity for an electron passing at $\rho$ = 5~nm from a nanoparticle excited with 800~nm light. For 20-100~nm particles, optimal coupling ($\beta_{opt}$) occurs at SEM energies between the vertical dotted lines. The value of $\beta_{opt}$ when $\rho<<\lambda$ follows a simple linear relationship, shown by the dashed line. (b) Plot of the optimal $\beta_{opt}$ for an optical nanostructure for a given thicknesses and excitation wavelength. Again, the white dotted lines divide zones corresponding to boundaries between LEEM, SEM, and TEM energies.}
    \label{fig:np-loss}
    \vspace{-3mm}
\end{figure}

We also explored if stronger interactions would apply to interactions with other nanoscale materials imaged at SEM energies. Specifically, we wanted to understand if aloof excitation of nanostructures would be improved as well. To do this we simulated 1-300~keV electrons interacting with the near-fields of optically-driven gold nanoparticles. We found that optimal coupling to the optical near-fields of the gold nanoparticles similarly occurred at SEM energies, which we illustrate in Fig. \ref{fig:np-loss}a. 

This maximal coupling can be understood intuitively by considering the velocity matching condition between the passing electron and the oscillating near fields. The electron gains (or loses) the most energy from the particle when half the cycle time of the optical excitation matches the transit time of the electron through the zero-crossings of the near fields. If the electron impact parameter is significantly smaller than the exciting wavelength, by symmetry this crossover extends from the nanoparticle at a 45$^\circ$ angle, and the length of travel is simply $d + 2\rho$, as illustrated in the inset of Fig. 3a. Combining this with the optical cycle time results in the following prediction of the optimal relativistic velocity, $\beta_{opt}$ given by
\begin{equation}
\label{eqn:couplFit}
    \beta_{opt} \approx \frac{ d + 2\rho}{\lambda/2},
\end{equation}
where $\lambda$ is the photon wavelength, $d$ is the nanoparticle diameter, and $\rho$ the impact parameter. For a 50~nm  sphere illuminated by 800~nm light and a $\rho = 5$~nm, we find that this maximum occurs at around an electron velocity of around 12.5\% the speed of light, remarkably close to the simulated value of 12.4\%. We verify this behavior for a variety of impact parameters and structures in the supplement, finding that it valid for $\rho < \lambda/20$

For visible and near-IR plasmonic processes, this then means that SEMs offer the optimal coupling for a variety of wavelengths and thicknesses, illustrated in Fig.  \ref{fig:np-loss}b. This can be thought of as the result of the wide dynamic range of slow electrons in this non-relativistic regime. Similar to how TEM's are excellent for coupling to low-refractive index materials ($n\leq3$) as previously shown \cite{feist_cavity-mediated_2022, feist_high-purity_2020}, the significant optical confinement of light due to surface plasmons means that much slower electrons will couple effectively to materials with optical confinement factors from 3-20 (matching the 1 to 30~keV range of SEMs), typical for many plasmonic materials.


While the study of 2D materials and optical nanostructures could greatly benefit from the improved coupling at SEM energies, the use of a modified SEM presents challenges not seen in TEM-based spectroscopy, including issues with imaging resolution, spectral stability, energy-dependent detector gain, and contamination, which limit the integration times and energy range of the instrument and are discussed in depth in the supplement. This meant that the signal-to-noise ratio (SNR) achieved in many of our measurements was poorer than expected, despite the fundamentally improved coupling. This is most clearly seen by the noisy spectra at lower beam energies plotted Fig. 2b-c, which seemingly contradicts our claim that SEM beam energies, and the higher $t/\lambda_i$ ratio associated with them, are the optimal range for such experiments. Fortunately these issues are not fundamental, and so if SEM-based spectroscopy instrumentation was developed to the sophistication and specifications of TEMs, this order-of-magnitude coupling improvement  would lead to equivalently improved SNRs.  

A particularly promising application for this technique is its use in ultrafast spectroscopy, which suffers from intrinsically poor SNR due to the low duty cycle of the optically-triggered electrons. The effective duty cycle of the microscope can be expressed as $D_{eff}=\sqrt{2}f\sigma_t$ , where $\tau=\sigma_t \sqrt{8\ln{2}}$  is the FWHM duration of the ultrafast pulse. If we assume relatively typical 200~fs pulses at an 800~kHz repetition rate \cite{feist_ultrafast_2017}, we would expect a roughly four-order-of-magnitude decrease in the brightness of the source.  For creation of a 10~nm probe, as is the case of this work, this would correspond to roughly 2 to 20 fA of effective current \cite{feist_ultrafast_2017}. Thus the two-order-of-magnitude improvement of coupling demonstrated in this work would go a long way in improving the range of measurable phenomenon with this approach, and allow for ultrafast techniques historically used on bulk materials to finally be applied to 2D films and optical nanostructures. 

In this work, we examined the use of lower velocity electrons for EELS of 2D materials and optical nanostructures, concluding that order-of magnitude performance improvements to the state-of-the-art are possible for this widely used technique. We demonstrated this for the graphene in a home-built electrostatic spectrometer integrated into a 1-30 keV SEM. We found that the scaling of coupling to surface plasmons in graphene versus acceleration energy follows the same power-law scaling as bulk materials, albeit with a significant offset, which we hypothesize is due to surface effects dominating over the bulk effects assumed in the model. We found that this offset factor is supported by trends in both LEEM and TEM experiments from the literature, and leads to optimal electron-material coupling at the energy range used. We then extended this model to other 2D materials, finding that their optimal coupling should also fall into this energy range.  Finally, we extend this coupling argument to PINEM experiments using optically-driven plasmonic nanostructures, showing that 1-30~keV electron energies lead to optimal coupling for a wide variety of wavelengths and structure scales. Based on our findings, we conclude that SEMs are potentially an ideal platform for next generation exploration of novel nanostructure and 2D material systems, especially in ultrafast electron pump-probe regimes.

\begin{acknowledgments}
We are grateful to Dr. Yugu Yang-Keathley, Dr. Matthew Yeung, Camron Blackburn, and Dr. Felix Ritzkowsky for helpful discussions on the design and construction of the spectrometer, as well as for reviewing this manuscript.  The construction of the electron spectrometer was supported by the U.S. Department of Energy, Office of Science, Office of Basic Energy Sciences, under Award Number DE-SC0022054. The calibration samples, electron-photon coupling analysis, and simulations were supported by the National Science Foundation under Grant No. 2110535 (MIT) and No. 2110556 (UC Davis).  

\end{acknowledgments}

 \bibliography{references2.bib}
 
\newpage

\appendix
\section{Methods}
\label{Sec:methods}

\subsection{Spectrometer Operation}
Our spectrometer is a custom-built electrostatic $127^\circ$  cylindrical analyzer, with a 70~mm radius and 5~mm plate separation. At the entrance aperture (located 2.5~mm above the plate start), we place a TEM grid for alignment. At the exit plane we place either a phosphor or a 5 to \SI{25}{\micro\meter} slit. In the case of the slit, either a electron sensitive photodiode (Opto Diode Corp AXUV20HS1) or Channeltron electron detector is used to amplify the beam, depending on if we are in a current-measuring or electron counting regime. For this work, we used the electron sensitive diode coupled to a low-noise amplifier (SRS570, Stanford Research Systems) and oscilloscope.

For 10 keV experiments, the upper and lower electrodes are biased to + 720 V and -720 V respectively. For other energies, the center pass energy is scaled linearly (so at 5 keV we require approximately 360~V). Realistically this value must be changed by about 5\% to account for day-to-day variations in the setup. 

In order to scan the energy of our beam, we apply a maximum 20 volt peak-to-peak triangle wave into the bottom plate of our spectrometer, via a bias tee, ranging from 10 - 1 kHz. Given the bias tee has a \SI{0.47}{\micro\farad} capacitance to the plates, which have approximately 10~pF of capacitance to ground, minimal distortion is expected. We have a \SI{10}{\mega\ohm} resistor between the spectrometer and source to protect the high voltage supply.

\begin{figure}
    \centering
    \includegraphics[width=1\linewidth]{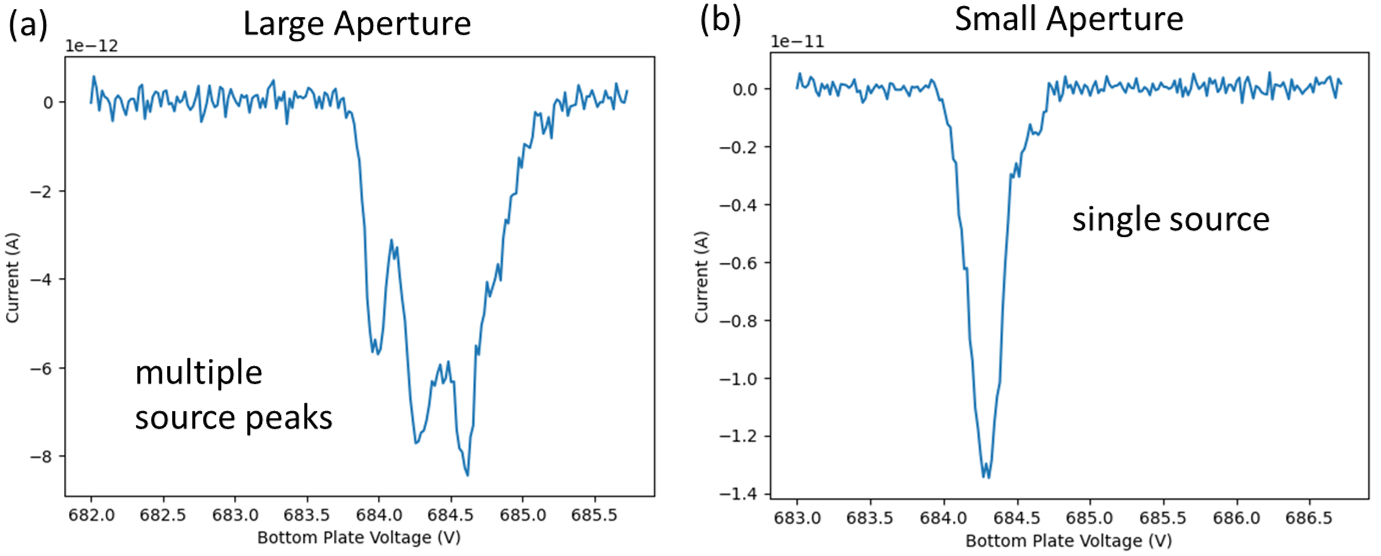}
    \caption{Effect of aperturing on the electron beam energy distribution. (a) Standard focusing alignment of the electron beam in the SEM leads to multiple peaks in the spectrometer, which we hypothesize is due to multiple facets of our electron source leading to a well-defined, incoherent probe, but multiple energy peaks. (b) If we aperture our beam further, we can isolate single electron peaks from the source and thus achieve high resolution imaging. In this case, the top spectrometer plate is held at 704~V, meaning the energy FWHM corresponds to 1.75 eV. After this step further alignment of the slit is required }
    \label{fig:appDemo}
\end{figure}

Spectrometer alignment occurs at 10 keV. First, we align the electron beam at high current to the column, and then mechanically translate the spectrometer to lie and this same optical axis. We then scan over the beam and center it. We iterate between rotating the slit with a mechanical feed through to minimize the width of the pass peak, and increasing the defocus on the condenser aperture to select a smaller portion of the beam. In normal SEM operation, we note that we see many peaks in the spectrum of the beam, seen in Fig. \ref{fig:appDemo}a. We hypothesize that this is due to the incoherent imaging mode of secondary-electron detection, in which many facets of the electron source can still result in a good-looking SEM micrograph. If we adjust the condenser while reading at the spectrum, we can filter out these orders down to a single, 0.7~eV peak, seen in Fig. \ref{fig:appDemo}b.

For such low energy experiments, drift is a major problem. In order to compensate for this, we begin our experiments by rapidly scanning our spectrometer at 1~kHz to sample any high frequency fluctuations. After these have been accounted for by removing noise sources (such as vibrations or feedback in our electromagnetic cancelling system (Spicer Consulting) or waiting for them to disappear, we reduce the scan speed to 60 Hz. The output spectrum is first amplified with a variable-gain, variable-bandwidth amplifier (SRS SR570) before being coupled to an oscilloscope (Keysight DSOX6002A). This rapid scanning is summarized in Fig.  \ref{fig:scanning_comp} below.

\begin{figure*}
    \centering
    \includegraphics[width=1\linewidth]{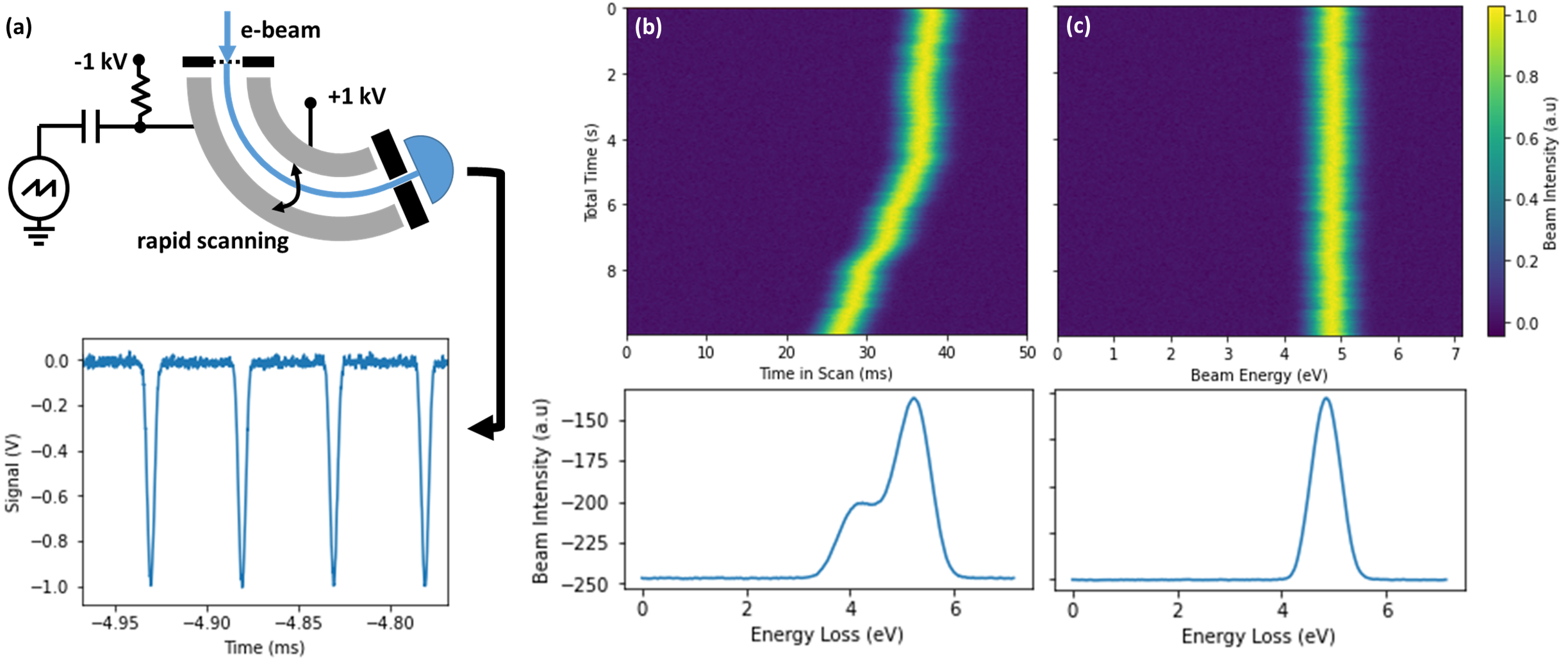}
    \caption{Fast scanning in our spectrometer. By rapidly sweeping the beam faster than instabilities in our system, we can compensate them out be measuring the ZLP position. (a) The spectrometer configuration, in which the bottom plate has a bias tee configuation to couple in a fast modulated trangle wave. This results and a repeated spectrum shown below for four periods. (b) The raw spectrum, showing drift in the peak position. If we naively integrate the signal along un-adjusted time bins, we can get the appearance of spurious sidebands, even in short integration windows of a few seconds. (c) Compensated signal, in which we find the ZLP peak, and shift our data on each row to this value. Note that this is not perfect and leads to a random spread of our beam. Because of the summing of these random errors our spectrum appears closer to a normal distribution rather than the expected Maxwell-Boltzman distribution given this energy.}\label{fig:scanning_comp}
\end{figure*}

We then image the zero loss peak. Starting with no aperture means that our maximum energy resolution is on order 35 eV, due to many incoherent electron source points on our emitter. Introducing a \SI{10}{\micro\meter} aperture and adjusting the condenser to achieve good SE imaging greatly ameliorates this, allowing for us to approach 5 eV resolution. At this point, we rotate the slit to account for rotation of the dispersed beam due to stray DC magnetic fields, until various source peaks become visible. We then select an individual source peak with the condenser and aperture, allowing for the peak 0.65 eV resolution observed. For spectroscopy of higher energy loss peaks, allowing multiple orders to be part of the ZLP can be helpful, as signal intensity can be greatly increased by allowing the resolution to worsen to around 2~eV.

\subsection{Sample Preparation}
Our starting samples were monolayer graphene on lacey carbon on gold TEM grids from Ted Pella (Prod. 21710). We then placed these onto a hot plate with a glass slide underneath, heated to \SI{100}{\celsius}. Gold nanoparticles of 40~nm in solution (Sigma-Aldrich 753637), diluted by a factor of 10:1 in deionized water was then drop-cast onto the TEM grids. These were used for alignment, and to try probing 2.3 eV plasmon sidebands. While such sidebands were observed on occasion, the large instability of our system (about 10~nm vibrations) prohibited the careful measurement of these features.

This sample was then loaded into our holder with silver paste, and placed into the input aperture plane of our spectrometer.

\subsection{Challenges of Low-Energy Operation}

In performing these experiments, we encountered several challenges unique to SEMs.  In the imaging plane, fast vibrations on the order of 20~nm, and slower drift on the order of 100~nm meant that spatial resolution was limited.  If the ZLP was too small ($t/\lambda_i > 1$), we could no longer align each peak and sum them to reduce noise, which limited the maximum values we could measure, especially for lower energy graphene measurements. Improved shielding and mechanical design of the stage and column, to the same degree as TEMs, should substantially improve this. If this were resolved, resolution of plateaus of varying 2D material thicknesses would be possible, greatly reducing uncertainty in our estimates.

Contamination was another challenge, which we attribute to the quality of an SEM vacuum, which is on order $\mathrm{10^{-6}}$ Torr relative to the $\mathrm{10^{-9}}$ Torr vacuum found in TEMs. If focused to under 50~nm we would deposit roughly a monolayer of contamination in 10~s, and so for graphene measurements we defocused the beam significantly, and varied beam placement between measurements. Regardless, this contamination limited our integration time on the sample. With a better vacuum, less contamination would mean longer sample integration times would be possible.

Variable gain in our detector was another problem. When using low-energy semiconducting electron detectors, the incident output current is proportional to the electron energy. This means that as we decreased the electron energy, while coupling became better the SNR actually got worse. This is also addressable, as if instead of a junction-based detector and amplifier we used an electron counter, the SNR would not decrease as dramatically with energy.

\section{Modeling}
\subsection{Inelastic Mean Free Path Calculations}
\label{Sec:theory-2D}
In EELS, electron-sample interactions are generally treated as random scattering events from the electron with the specimen of interest. These interactions can be due to a wide variety of processes, such as phonons scattering, plasmon interactions, core-electron excitations, and contribute to a general loss of signal intensity in the zero-loss peak ($I_\text{ZLP}$) compared to the initial intensity ($I_0$). Averaged over many electrons, this loss follows an exponential decay, given by

\begin{equation}
    \frac{I_\text{ZLP}(t)}{I_0} = e^{-t/\lambda_i}.
\end{equation}

For the first plasmon peak, which is the dominant term we will be analyzing, the integrated area is given by:

\begin{equation}
    \frac{I_{p1}(t)}{I_0} = \frac{t}{\lambda_i}e^{-t/\lambda_i}.
\end{equation} \label{eqn:plasmon1}

The intensity of other loss features such as core-losses (seen as sharp edges) and interband excitations follow similar statistics, with their overall strength dictated by $\lambda_i$, and so the analysis presented here is relevant for those features as well.

We begin our estimate of 2D material inelastic cross sections for graphite and hBN, we take the inelastic mean free path as calculated using the full Penn Algorithm from optical data \cite{jablonski_calculations_2023}. For all other materials, we use the modified Bethe equation, which is more approximate, but considered to be largely valid for the range of 100 eV to 200 keV. This equation is given by

\begin{equation}
    \lambda_i(E_0)=\frac{\alpha(E_0) E_0}{E_p^2[\beta[\ln (E_0)-(C / E_0)+(D / E^2)]}
\end{equation} \label{eqn:bethe}

where $\lambda_i$ is given in nm, and

\begin{equation}
\alpha(E)=\frac{1+E /\left(2 m_e c^2\right)}{\left[1+E /\left(m_e c^2\right)\right]^2} \approx \frac{1+E / 1021999.8}{(1+E / 510998.9)^2}
\end{equation}

is the the relativistic factor,

\begin{equation}
E_p=28.816\left(\frac{N_v \cdot \rho}{M}\right)^{0.5}
\end{equation}

the plasmon energy, and the constants in eqn. \ref{eqn:bethe} defined by

\begin{equation}
\begin{aligned}
&\gamma=0.191 \rho^{-0.5}\left(\mathrm{eV}^{-1}\right)\\
&\mathrm{C}=19.7-9.1 U\left(\mathrm{~nm}^{-1}\right) \text {, }\\
&D=534-208 U\left(\mathrm{eV} \mathrm{nm}^{-1}\right) \text {, }\\
&U=\frac{N_{\mathrm{v}} \rho}{M}=\left(E_{\mathrm{p}} / 28.816\right)^2
\end{aligned}
\end{equation}

are the relevant parameters. These are derived from the material density, $\rho$, the number of valence electrons, $N_v$, and the molar mass of the constituent atoms/molecules. Various examples are plotted below

In the given reference, this form matches the results from the optical Penn algorithm to within 10\% root mean square (RMS). However it is well known for 2D materials this estimate is an over-estimate of the mean free path. In the case of hBN and graphene, we find that this Bethe approximation estimates a mean free path approximately 20\% RMS larger than than what we observe over the full energy range. We expect this to also occur for the other 2D materials we analyzed, which we have no optical data for.

We plot the various 2D materials referenced in Fig. 2 of the main text below, to give a sense of how we would expect other 2D materials to behave in the TEM. The code used to generate these plots, as well as the reference values used for them, are provided in our supplemental materials.

\begin{figure}
    \centering
    \includegraphics[width=1\linewidth]{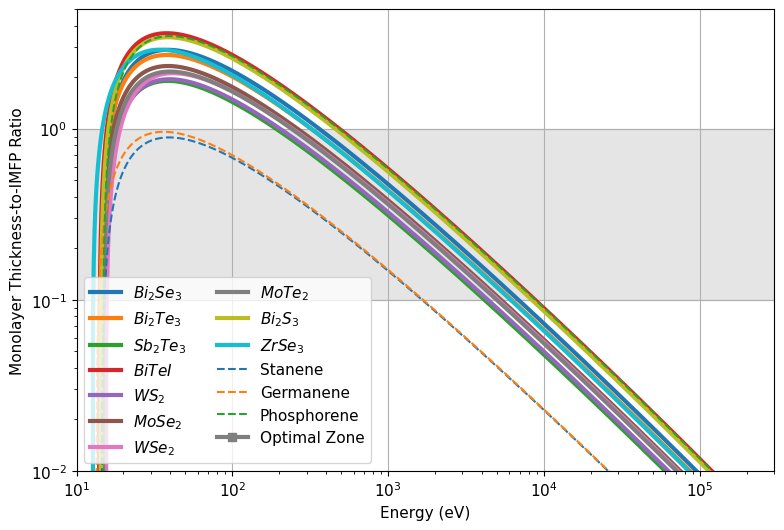}
  \caption{Various predicted 2D material curves matching the grey dotted "Var. 2D Mat." curves of Fig. 2 of the main text.}
    \label{fig:enter-label}
\end{figure}
The bigger issue is that this analysis assumes bulk plasmon coupling as the dominant term to estimate the IMFP \cite{penn_electron_1987}. This is clearly not valid in the case of 2D materials. Other studies have addressed this issue and created better models in low-energy regimes (\cite{geelen_nonuniversal_2019}), though to our knowledge none exists for the 1-300 keV range we aimed to compare. However given that the general trend of lower electron energies having stronger coupling from TEM to SEM energies, our own observations, and experiments from the literature, we found that this simple analysis is sufficient to indicate the overall idea that optimal EELS should occur at significantly lower energies that previously utilized in TEMs.

\subsection{Optical Nanostructure Simulations}
We use the frequency-domain electromagnetic waves module in COMSOL to simulate the near-field profiles of an illuminated gold nanoparticle. This tool uses the Finite Element Method to numerically solve the time-harmonic wave equation for the electric field. The simulation domain consists of a gold \cite{johnson_optical_1972} sphere with a varying diameter surrounded by a spherical shell of vacuum, surrounded by a perfectly matched layer. The sphere is illuminated by a \SI{800}{\nano\meter} light propagating in the $z$ direction and polarized in the $x$ direction. The theoretical electron path is parallel to the $x$-axis, at various impact parameters offset from the sphere in the $z$ direction. The PINEM coupling efficiency $g$ is calculated for each electron path corresponding to the sampled sphere diameters and impact parameters, and for a range of electron velocities:
\begin{equation}
    g=\frac{q}{\hbar\omega}\left|\int_L\tilde{E}_z \exp{\left(i\frac{\omega}{v}z\right)}dz \right|
\end{equation}
The electron path is indicated by $L$, and $v$ is the electron velocity. For each sphere diameter and impact parameter, there is an optimal electron energy to maximize the coupling efficiency, as is illustrated in Fig. 7. Note, $g$ is normalized per each diameter and impact parameter; the absolute coupling efficiency decreases at farther impact parameters. The general trend for the normalized coupling efficiency shows that smaller nanoparticles are more favorably coupled to by lower energy electrons. At farther impact parameters, the optimal coupling point shifts to higher electron energies. 

\begin{figure*}
    \centering
    \includegraphics[width=1\linewidth]{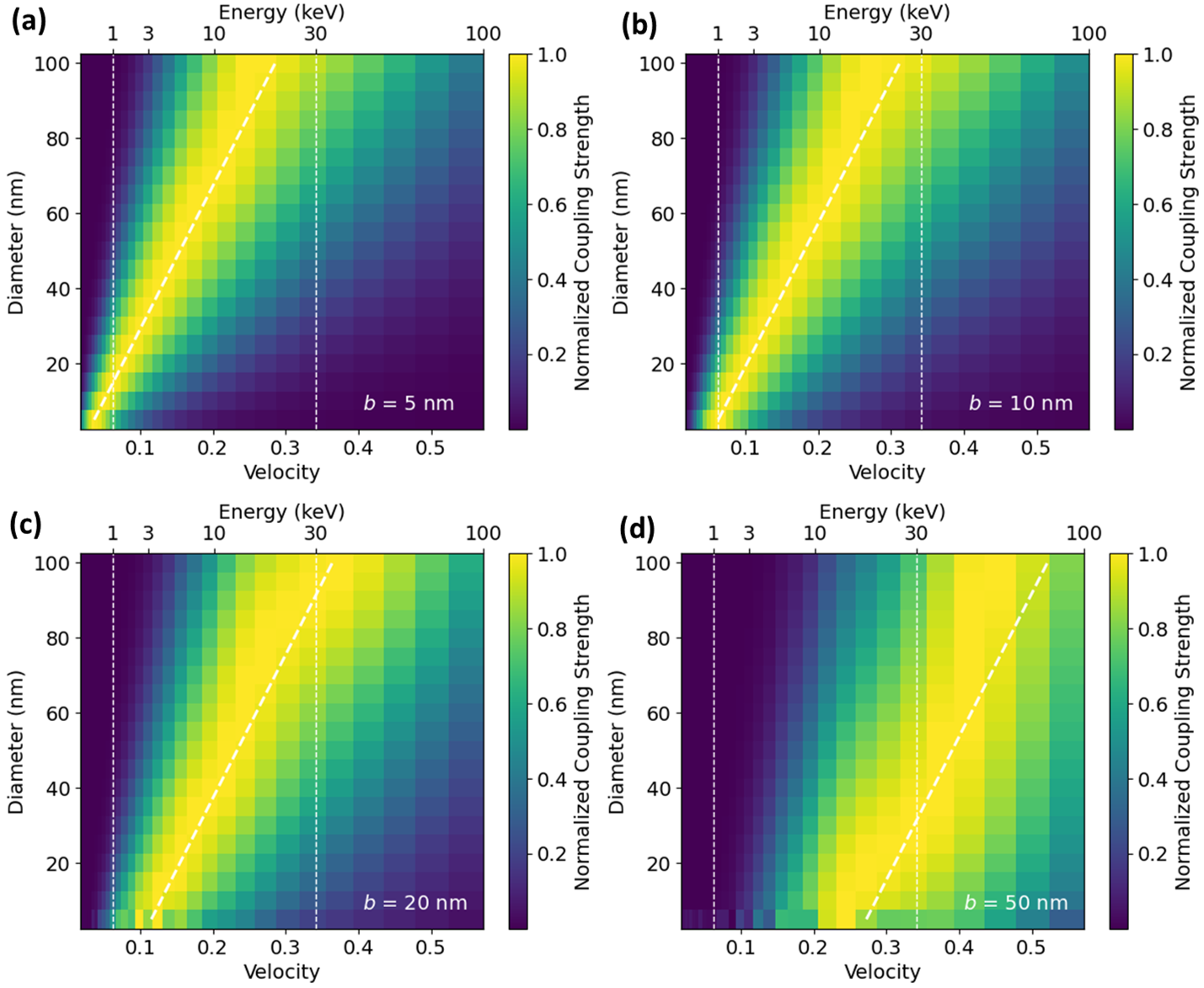}
    \caption{Plots of the simulated coupling efficiency versus electron velocity for an electron passing near a nanosphere at impact parameters of (a) 5 nm, (b) 10 nm, (c); 20 nm, and (d) 50nm. The plots are overlaid with lines indicating the optimal coupling velocity derived from the simple phase matching condition. As the impact parameter increases, this becomes less valid}
\label{fig:nanosphere_sup}
\end{figure*}


\end{document}